\documentclass[aps,prl,twocolumn,showpacs,preprintnumbers,amsmath,amssymb]{revtex4}

\usepackage{graphicx}% Include figure files
\usepackage{dcolumn}% Align table columns on decimal point
\usepackage{bm}% bold math
\usepackage[usenames,dvipsnames]{color}
\usepackage[colorlinks,linkcolor=blue,citecolor=blue]{hyperref}

\begin{document}

\title{Crowd synchrony and quorum sensing in delay-coupled lasers}

\author{Jordi Zamora-Munt}
\affiliation{Departament de F\'isica i Enginyeria Nuclear, Universitat Polit\`ecnica de Catalunya,
Edifici GAIA, Rambla de Sant Nebridi s/n, 08222 Terrassa, Barcelona, Spain}

\author{C. Masoller}
\affiliation{Departament de F\'isica i Enginyeria Nuclear, Universitat Polit\`ecnica de Catalunya,
Edifici GAIA, Rambla de Sant Nebridi s/n, 08222 Terrassa, Barcelona, Spain}

\author{Jordi Garcia-Ojalvo}
\affiliation{Departament de F\'isica i Enginyeria Nuclear, Universitat Polit\`ecnica de Catalunya,
Edifici GAIA, Rambla de Sant Nebridi s/n, 08222 Terrassa, Barcelona, Spain}

\author{Rajarshi Roy}
\affiliation{Institute for Research 
in Electronics and Applied Physics, Department of Physics, and
Institute for Physical Science and Technology, University of Maryland, College Park, MD 20742, USA}

\date{\today}

\begin{abstract}
Crowd synchrony and quorum sensing arise when a large number of dynamical elements communicate with each other via
a common information pool. Previous evidence in different fields, including chemistry, biology and civil
engineering, has shown that this type of coupling leads to synchronization, when coupling is
instantaneous and the number
of coupled elements is large enough. Here we consider a situation in which the transmission of
information between the
system components and the coupling pool is not instantaneous. To that end,
we model a system of semiconductor
lasers optically coupled to a central laser with a delay. Our results
show that, even though the lasers are non-identical due to their distinct optical frequencies, zero-lag
synchronization arises. By changing a system parameter, we can switch between two
different types of synchronization transition. The dependence of the transition with respect to 
the delay-coupling parameters is studied.
\end{abstract}


\maketitle

Many situations in nature involve systems of multiple dynamical elements that interact with each other
through a common medium. Examples include pendulum clocks mounted on the same
wooden beam \cite{HuygensClk}, cellular populations communicating via small molecules that freely
diffuse among the cells \cite{mcmillen,repress}, and longitudinal modes of a laser connected through
saturation of the common amplifying medium \cite{tang,wiesenfeld}. In most cases, the coupling surrounding
leads to a synchronous behavior
between all the coupled elements, but no general framework for such {\em crowd synchrony} has
been developed yet.

One of the features that determines whether a coupling medium is able to elicit synchronization is
the number of elements that are connected through it. One situation where this fact has been established
is that of the Millennium Bridge in London. Two days after its opening in 2000, the pedestrian bridge had
to be closed because of excessive wobbling. Subsequent investigations \cite{TestBridge} revealed that the pedestrians, who initially walked with different frequencies and phases, fell into a synchronous pacing
when the number of people on the bridge was large enough.
That effect was modeled and understood in terms of the synchronization
of simple oscillators \cite{NatureBridge,eckhardt}. In a biological context, collective glycolytic oscillations in yeast
cells have also been seen to arise as a result of large enough cell densities \cite{aldridge,ProcMonte}. In that
case, and in contrast with the Millenium bridge, oscillations of individual cells only occur above a critical number of them: the coupling
induces both the oscillations and the synchronization among them. A similar situation was recently
reported on a chemical oscillator system, formed by catalytic particles suspended in a catalyst-free
Belousov-Zhabotinsky medium. This system also shows a transition to synchronized oscillations
above a certain density of catalytic particles \cite{ScienceChemyOscil}. Furthermore, depending on the
coupling strength, the synchronization appears either progressively or suddenly. The question of
how generic this behavior is remains open.

Studies of crowd synchrony to date have considered the coupling with the medium to be
instantaneous \cite{NatureBridge,eckhardt,aldridge,ProcMonte,ScienceChemyOscil}. This naturally results in synchronous behavior with zero time lag between any pair
of elements in the system. However, in many situations the transmission of the coupling signal
takes an amount of time that is non-negligible with respect to the characteristic time scales of the
system components. This is the case e.g. in systems of technological importance such as
optically coupled semiconductor lasers. When these devices are separated distances on the order of centimeters, they are subject to coupling 
delays on the order of the characteristic time scales of solitary lasers (shorter than nanoseconds). 
In recent years much effort has been devoted to understand the synchronization of semiconductor lasers.
This is important for technological reasons, i.e., to achieve large output powers, but also for increasing
our knowledge of how generic dynamical systems synchronize. Semiconductor lasers are low-cost, versatile,
and many of the commercial lasers are well-described theoretically. They also show a wide variety of
nonlinear dynamical behavior, both as single elements with external influences and as part of laser networks; examples
include low-frequency fluctuations \cite{FischerLFF,heil}, chaos \cite{wieczorek1}, excitability
\cite{giudici,mulet,wieczorek2}, and nontrivial synchronization phenomena \cite{masoller,episodic}.
However, most studies of coupled lasers so far have considered a small number of
elements. Thus, how to achieve synchronization for a large number of coupled non-identical lasers
is still an open question.

In this Letter we show that a collection of $M$ semiconductor lasers coupled
through a central laser exhibits
zero-lag crowd synchronization. This setup is a generalization of the case of two identical oscillators
coupled through a third central element \cite{winful,terry,alan,FischerZeroLag,iacyel,zhou}.
Isochronal synchronization is relevant in both technological
\cite{vicenteOL} and biological \cite{vicentePNAS} contexts. Here the
central laser operates in a passive regime (below threshold), and plays the role of a coupling
medium analogous to the bridge structure in pedestrian synchronization \cite{NatureBridge},
and to the reaction medium in chemical synchronization \cite{ScienceChemyOscil}.
Our results show that the general properties of both the crowd synchrony and the
quorum-sensing transition are readily reproduced with this setup.

The equations describing the slow envelope of the complex electric field $E$ and the carrier density $N$
for the $M$ lasers are \cite{FischerZeroLag}
\begin{eqnarray}
\nonumber \dot{E}_{j}&=&i\omega_{j}E_{j}+\gamma\left(1+i\alpha\right)\left(G_{j}-1\right)E_{j}+\\
&&\kappa E_{H}\left(t-\tau\right)e^{-i\omega_{0}\tau}+\sqrt{D}\xi_{j}(t)
\\
\nonumber \dot{E}_{H}&=&i\omega_H E_{H}+\gamma\left(1+i\alpha\right)\left(G_{H}-1\right)E_{H}+\\
&&\kappa \overset{M}{\underset{j=1}{\sum}} E_{j}\left(t-\tau\right)e^{-i\omega_ { 0} \tau}+\sqrt{D}\xi_{H}(t)
\\
\dot{N}_{j,H}&=&\gamma_{e}\left(\mu_{j,H}-N_{j,H}-G_{j,H}\left|E_{j,H}\right|^{2}\right)\,,
\end{eqnarray}
where $G_{j,H}=N_{j,H}/(1+\epsilon\left|E_{j,H}\right|^{2})$, and the subscripts $H$ and $j$ denote the central (hub) laser and $j$th outer (star) laser, respectively.
The field and carrier decay rates are $\gamma$ and $\gamma_{e}$, respectively, $\alpha$ is the linewidth enhancement factor, $\epsilon$ is the gain saturation, $\omega_{0}$ is the optical frequency and $\omega_H$ and $\omega_j$ are the detunings of the hub and the star lasers with respect to the reference frequency $\omega_{0}$. The coupling is characterized by its strength $\kappa$ and delay $\tau$, both of which are assumed equal for all lasers.
$\mu_j=\mu$ and $\mu_H$ are the pump currents of the star and hub lasers, respectively.
Finally, $\xi_{j}(t)$ and $\xi_{H}(t)$ are uncorrelated complex Gaussian white noises,
with $D$ being the noise strength. 
The model was integrated with the stochastic Heun algorithm with a time step of 0.8~ps, using parameter
values typical for semiconductor lasers: $\gamma=300$~ns$^{-1}$, $\gamma_{e}=1$~ns$^{-1}$, $\alpha=3$, 
$D=10^{-5}$~ns$^{-1}$, $\omega_{0}=2\pi c/\lambda$ (where $c$ is the speed of light
and $\lambda=654$~nm). $\omega_H=0$ without loss of generality, and $\omega_j$ is chosen from a Gaussian distribution with zero mean and standard deviation $\sigma$.
In what follows we neglect nonlinear gain saturation ($\epsilon=0$), since it does not
affect the results obtained (not shown).
Unless otherwise stated $\sigma=20\pi$~rad/ns, corresponding to $\Delta \lambda=$0.014~nm,
and $\tau=5$~ns.

\begin{figure}[htb]
\includegraphics[width=8cm]{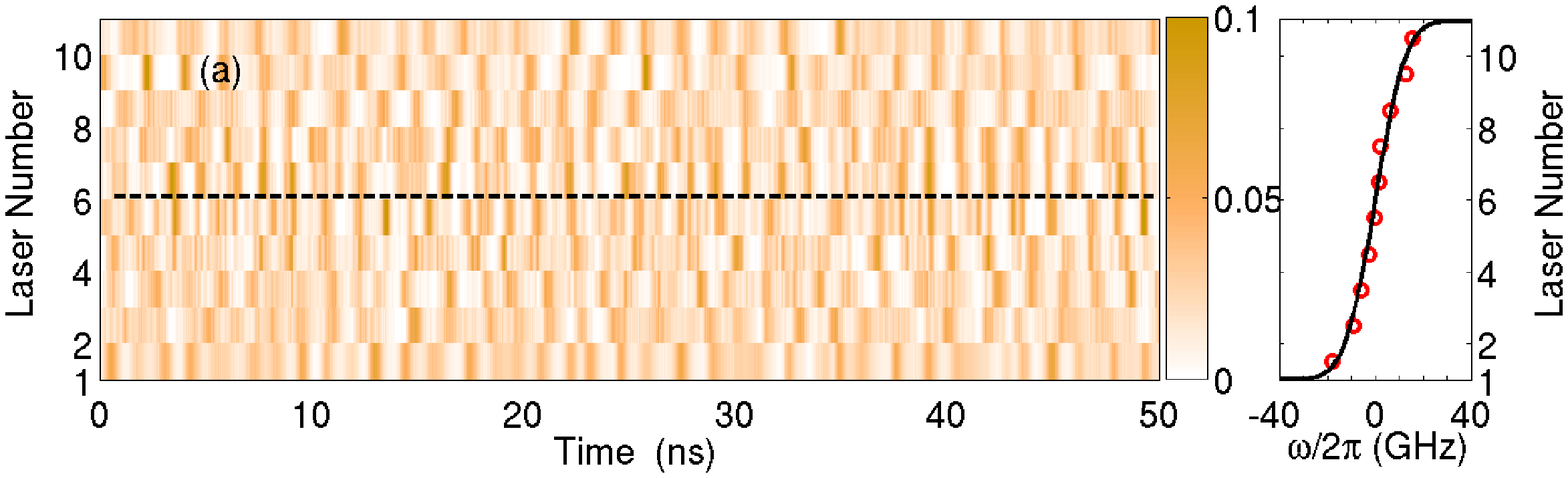}
\includegraphics[width=8.1cm]{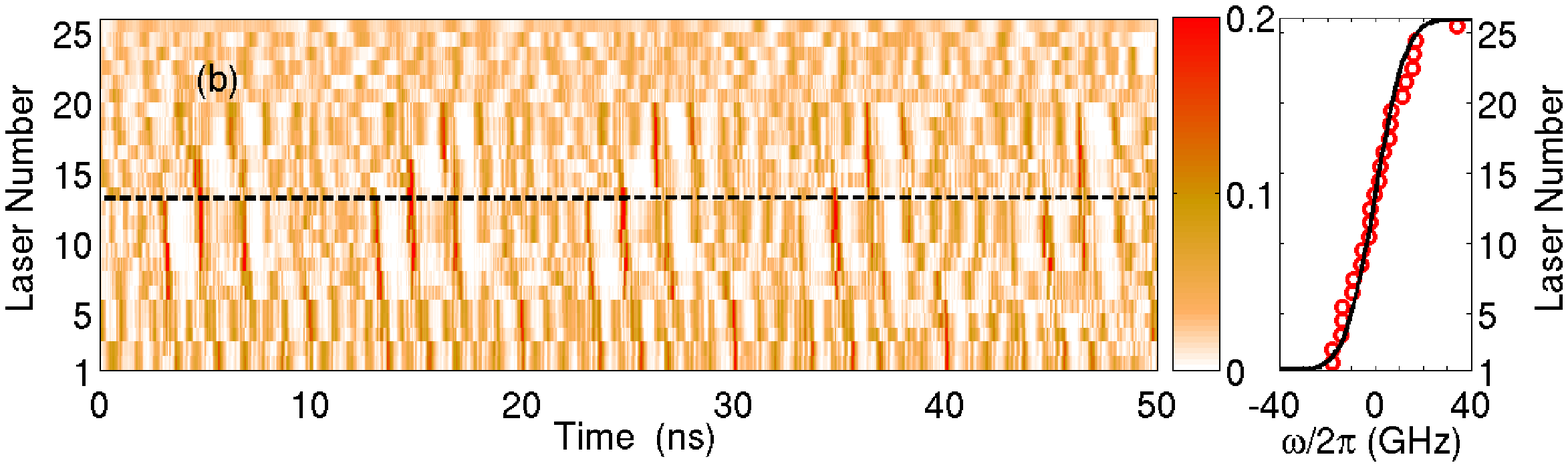}
\includegraphics[width=8.1cm]{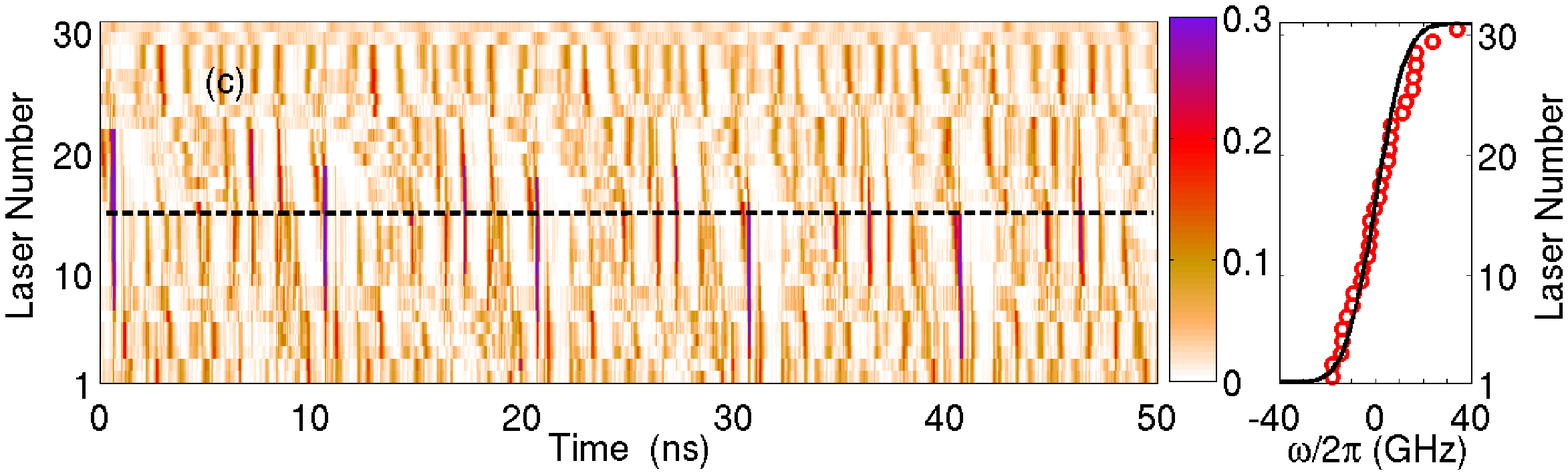}
\includegraphics[width=8.1cm]{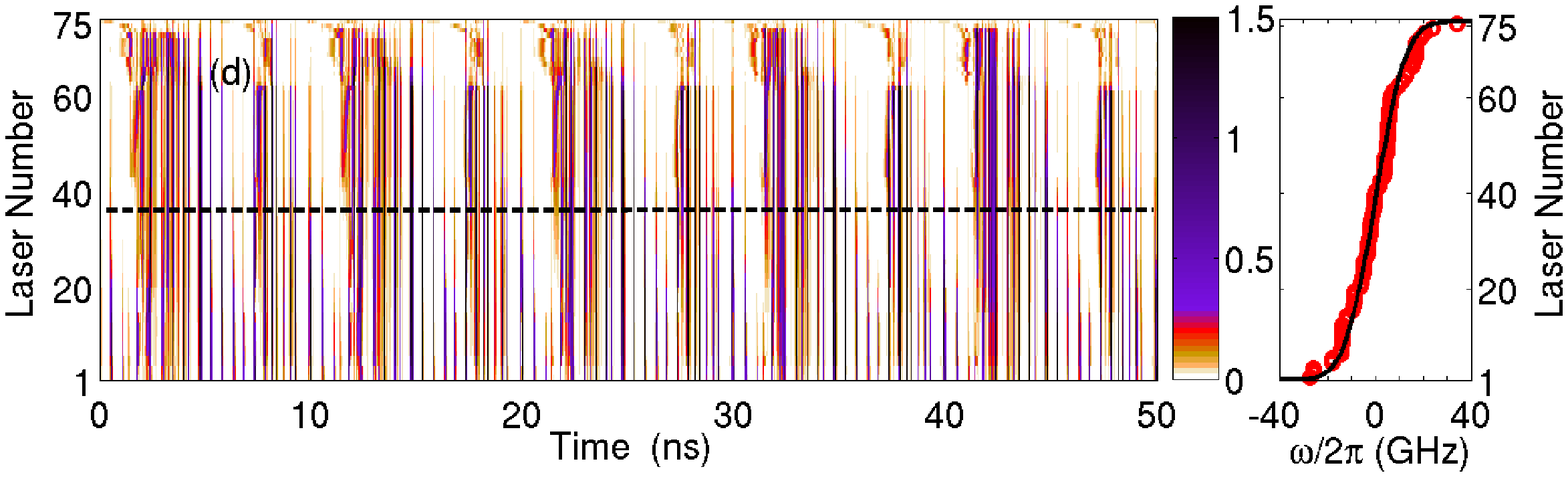}
\caption{\label{fig:1} (Color online) Synchronization of star-coupled semiconductor lasers for increasing number of elements, $M$. The color coding shows the intensity for each star laser as a function of time. In the vertical axis the lasers are sorted by their solitary frequency, $\omega_j$, with number 1 corresponding to the most negative  detuning. The black dashed line shows the laser for which $\omega=0$. (a) $M=10$, (b) $M=25$, (c) $M=30$ and (d) $M=75$. The right column shows the frequency $\omega_j/2\pi$ of the lasers (dots), in relation with the normailized cumulative Gaussian distribution (solid line). The pump currents are $\mu=$ 1.02 and $\mu_H=$ 0.4, and the coupling strength $\kappa=$ 30 ns$^{-1}$.}
\end{figure}

Figure \ref{fig:1} shows the stationary emitted intensity for varying number of star lasers. For small $M$ [Fig. \ref{fig:1}(a)] the lasers oscillate independently. By increasing $M$, synchronized emission at near zero-lag occurs for lasers with similar frequencies, forming clusters with similar dynamics as shown in Fig. \ref{fig:1}(b). The number of synchronized lasers in those clusters grows as $M$ increases [Fig.~\ref{fig:1}(c)], with an emission characterized by short pulses of irregular amplitudes with a repetition period around $2\tau$. Those characteristics become more evident for large $M$, where almost all the lasers emit synchronously at zero lag [Fig. \ref{fig:1}(d)], with emission pulses taking place simultaneously in most of the lasers. This situation is reached provided the pump current of the hub laser is set below the solitary lasing threshold, i.e. $\mu_H<\mu_{th}=1$ \cite{PumpAbove}.

\begin{figure*}[htb]
\includegraphics[width=17.5cm]{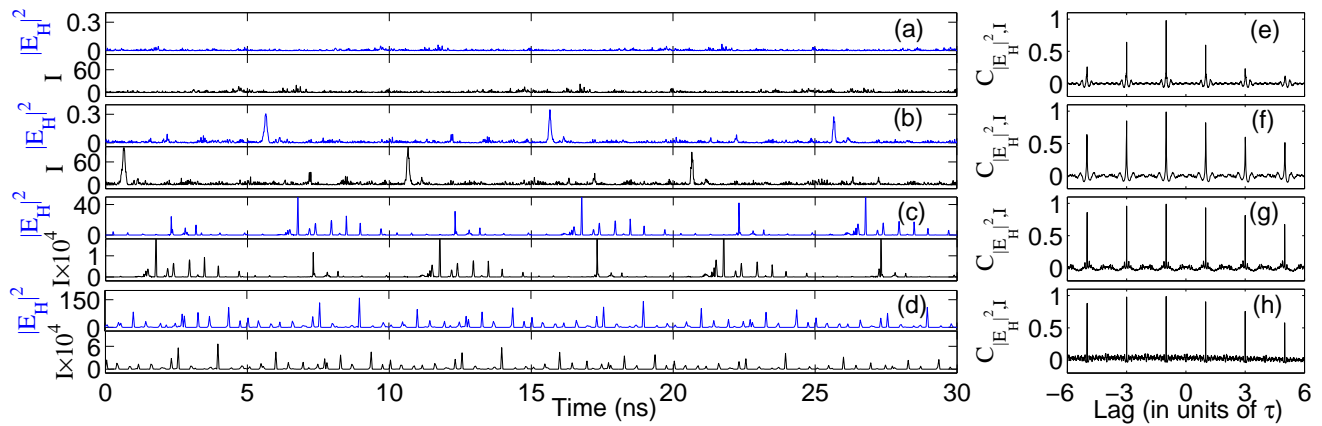}
\caption{\label{fig:1b} (Color online) (a-d) Time trace for the intensity of the hub laser, $|E_H|^2$ (top trace, blue), and for the coherent intensity, $I$ (bottom trace, black). (e-h) Cross-correlation function between $|E_H|^2$ and $I$. The number of lasers is $M=25$ (a,e), $M=30$ (b,f), $M=75$ (c,g) and $M=100$ (d,h). The parameters are the same as in Fig.~\ref{fig:1}.}
\end{figure*}

In order to quantify the level of zero-lag synchronization, we calculated the total coherent intensity of the star lasers as $I=\left|\overset{M}{\underset{j=1}{\sum}}E_{j}\left(t\right)\right|^{2}$. Figure~\ref{fig:1b} compares the dynamics of this quantity with that of the intensity $|E_H|^2$ of the hub for increasing number of lasers. For increasing values of $M$, Figs.~\ref{fig:1b}(a-d) shows the emergence of large intensity pulses in the total coherent intensity, corresponding to strongly synchronized activity in Figs.~\ref{fig:1b}(b-d). The hub laser reproduces these dynamics after a time $\tau$. This is reflected in a large peak at time $-\tau$ in the cross-correlation function between $|E_H|^2$ and $I$, shown in Figs.~\ref{fig:1b}(e-h). Thus, the hub laser lags behind the star lasers in the synchronized state.
 
To investigate the transition to the synchronized state, we use as order parameter the time-averaged total coherent intensity of the star lasers $\left\langle I\right\rangle$, where $\left\langle\cdot\right\rangle$ is the average over a time window of length $T= 4\;\mu$s, calculated in the stationary state. In the absence of synchronization
$\left\langle I \right\rangle$ grows linearly with $M$, while when synchronization arises this linear
dependence is lost.
\begin{figure}[htb]
\includegraphics[width=2.95cm]{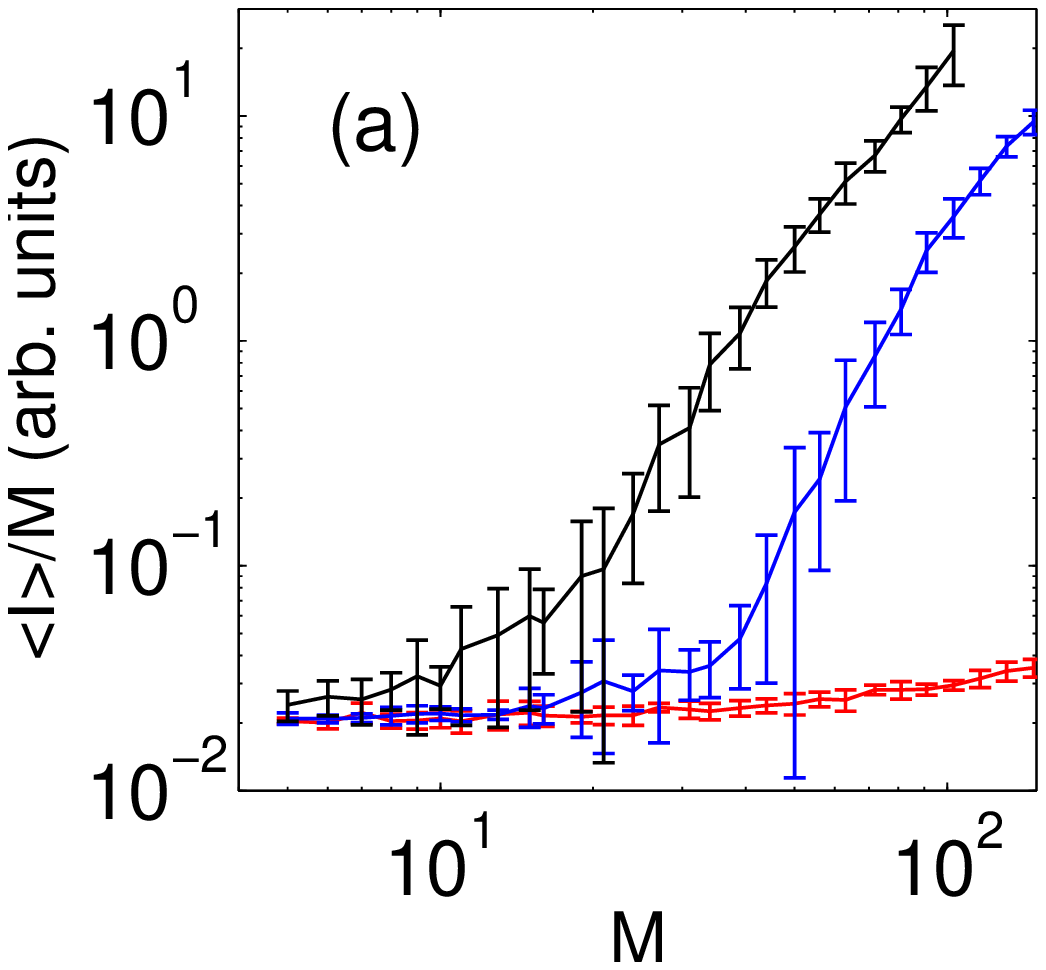}
\includegraphics[width=2.65cm] {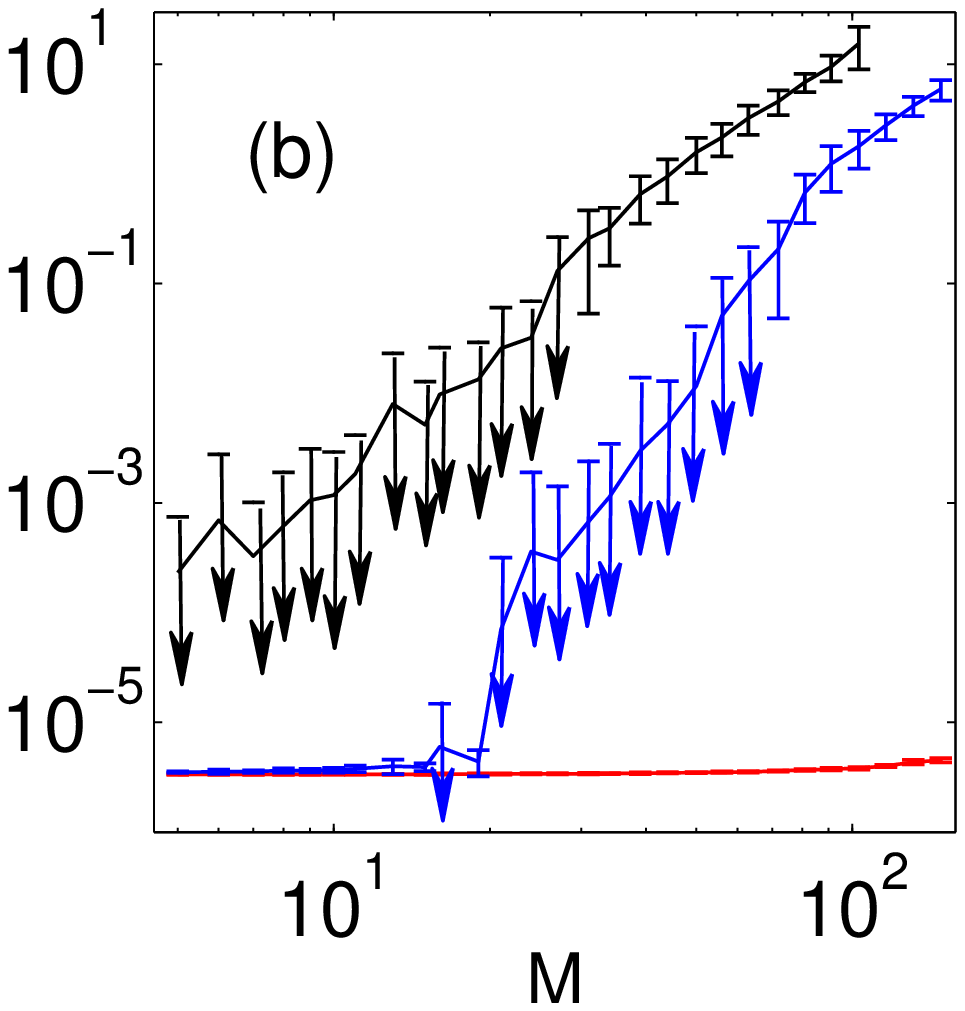}
\includegraphics[width=2.65cm] {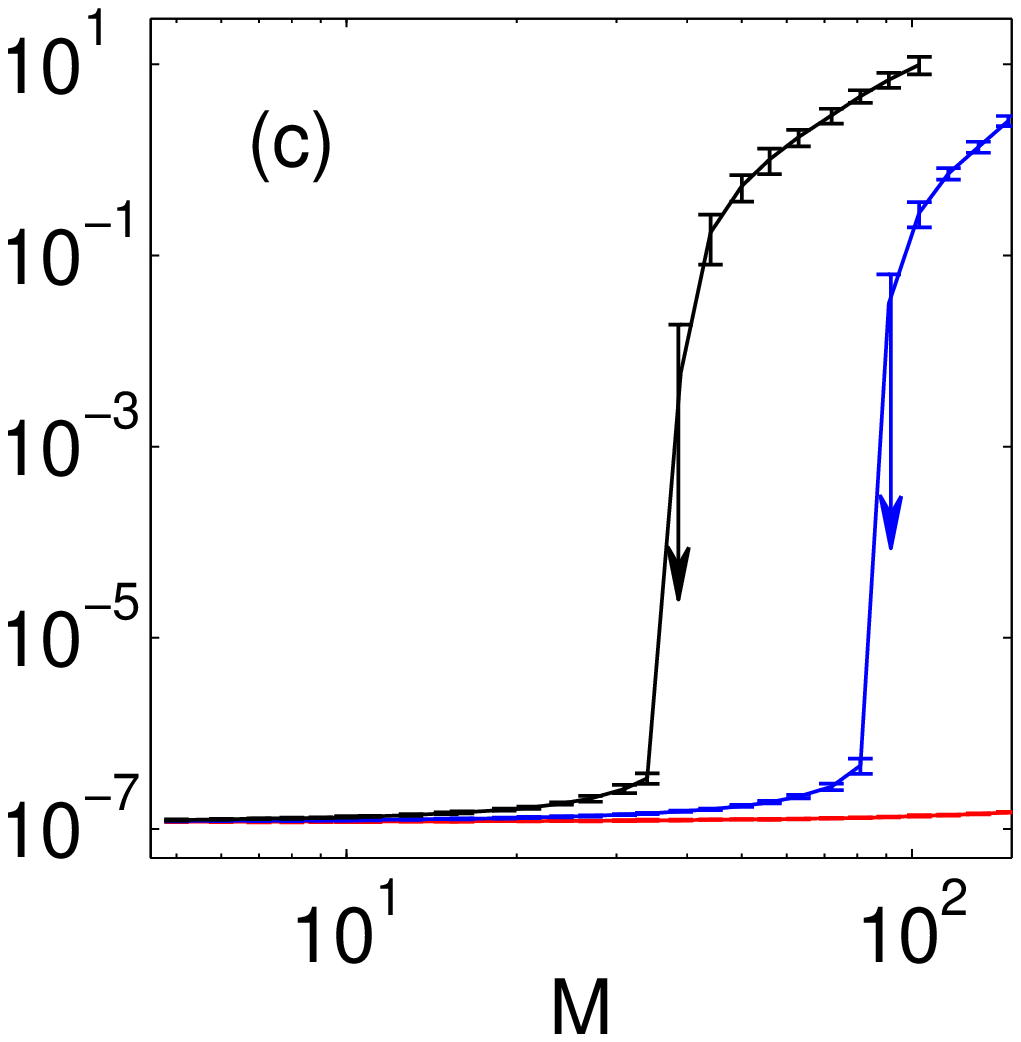}
\caption{\label{fig:2} (Color online) Ratio between the averaged coherent intensity $\left\langle I \right\rangle$
and the number of star lasers $M$, as a function of $M$ itself and for different coupling strengths: $\kappa=$ 10 ns$^{-1}$ (red), $\kappa=$ 20 ns$^{-1}$ (blue) and $\kappa=$ 30 ns$^{-1}$ (black). (a) $\mu=$ 1.02, $\mu_H=$ 0.4. (b) $\mu=$ 0.99, $\mu_H=$ 0.4. (c) $\mu=$ 0.7, $\mu_H=$ 0.4. Each point is averaged over 10 to 40 different initial conditions and detuning frequencies. The arrows mark errorbars out of the axis limits.}
\end{figure}
Figure~\ref{fig:2}(a) shows the average coherent intensity as a function of the number of star lasers for different coupling strengths and pump currents. When the star lasers are pumped above the solitary threshold
and for small coupling, $\left\langle I \right\rangle/M$
is approximately constant, corresponding to the case of the absence of synchronization. For moderate values of $\kappa$, on the other hand, the system becomes gradually synchronized as $M$ increases. The transition to synchronization occurs for a critical number of coupled lasers $M_c$, which
we quantify as the number of lasers for which the growth rate of $\left\langle I \right\rangle$ with $M$ changes
abruptly. For even larger $\kappa$ the critical number of lasers needed for synchronization decreases.

The qualitative character of the synchronization transition can be changed by tuning the pump current $\mu$
of the star lasers below the laser threshold. When $\mu$ is well below threshold [Fig.~\ref{fig:2}(c)],
the transition to synchronization is very sharp, in contrast with Fig.~\ref{fig:2}(a) above, provided coupling is large enough.
Note that in this case both the star and hub lasers are pumped below their solitary threshold, but are effectively above threshold due to coupling, and they turn on due to their spontaneous emission. The transition takes place when the star lasers are pumped at their solitary threshold [Fig.~\ref{fig:2}(b)], which shows the beginning of a sharp transition for intermediate $\kappa$ (blue line online) when the star lasers are pumped only slightly below threshold. We also note that this behavior requires that the hub laser be pumped below threshold, i.e. $\mu_H<\mu_{th}=1$,
otherwise the critical behavior is lost.

\begin{figure}[htb]
\includegraphics[width=5.4cm] {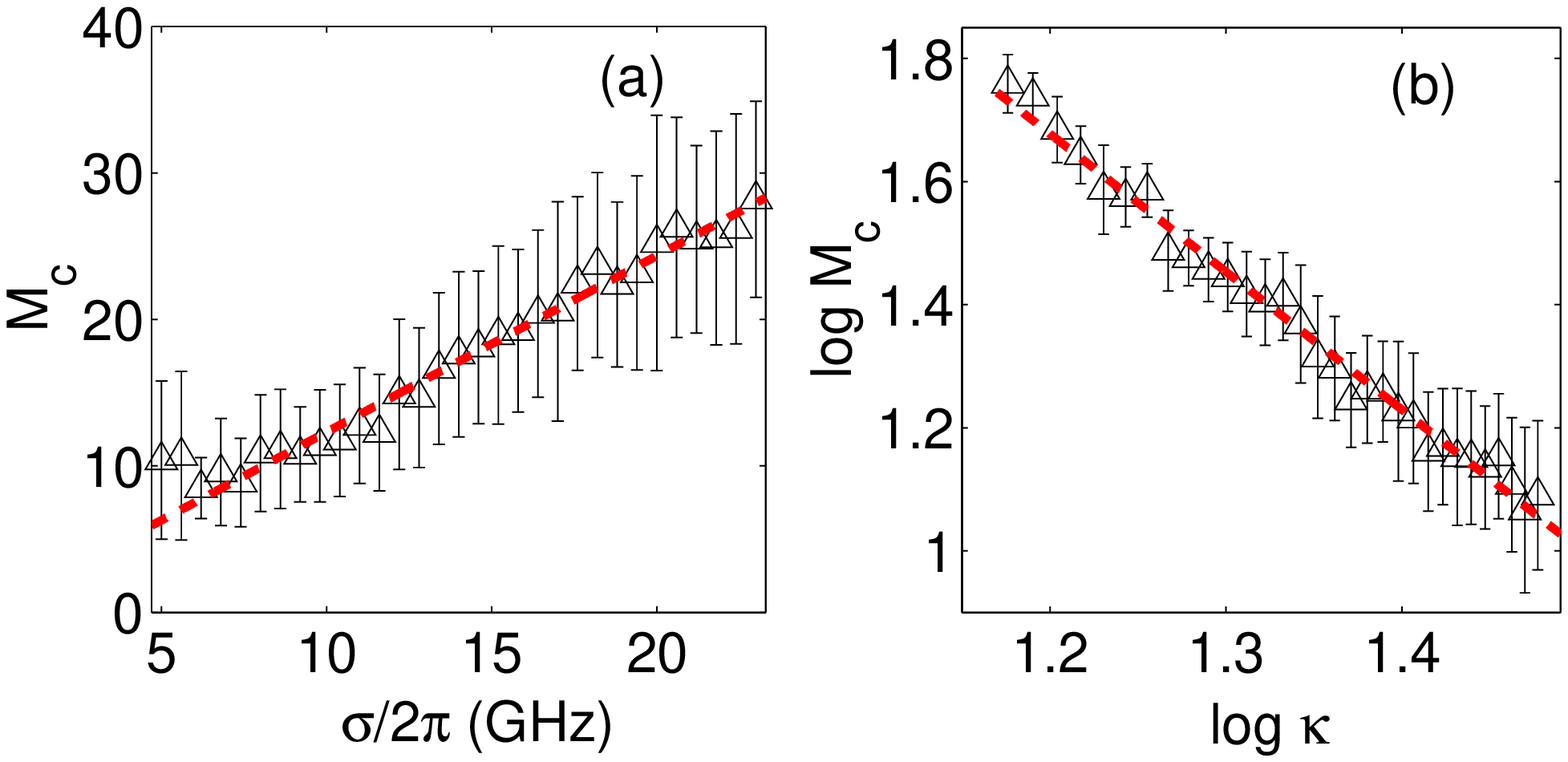}
\includegraphics[width=3cm] {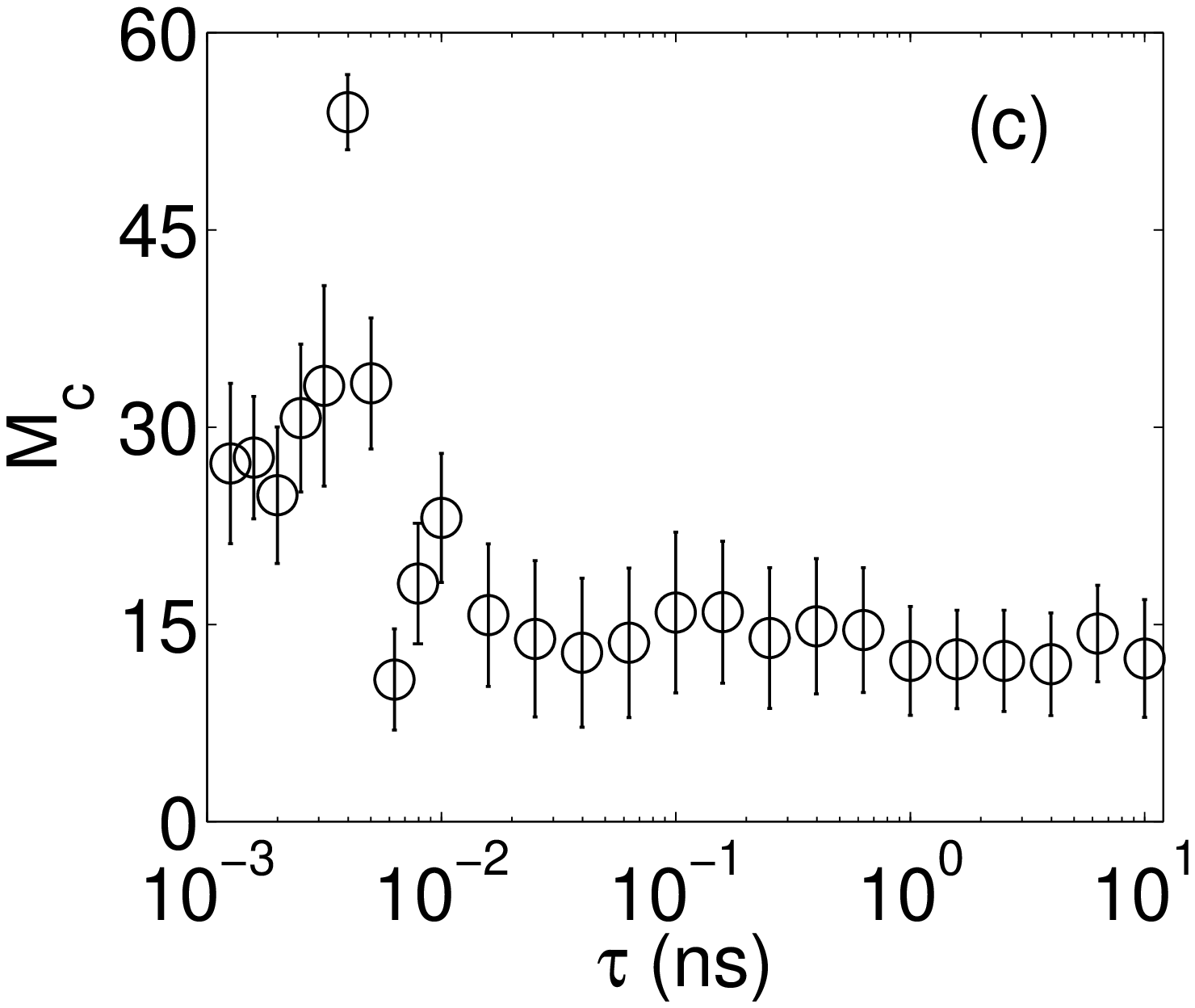}
\caption{\label{fig:3} (Color online) Critical number of lasers, $M_c$, as a function of different system parameters, for pump currents $\mu=$ 1.02, $\mu_H=$ 0.4. (a) $M_c$ as a function of the width of the frequency distribution, $\sigma$. The dashed line shows a linear fit for $\sigma/2\pi >6$ GHz ($M_c\propto1.2\sigma$, $r^2=0.982$). Other parameters are $\kappa=$ 30 ns$^{-1}$ and $\tau=$ 5 ns. (b) Doubly logarithmic plot of $M_c$ as a function of the coupling strength $\kappa$. The dashed line shows a power-law fit of the data ($M_c\propto1/\kappa^{2.2}$, $r^2=0.987$). Other parameters are $\sigma=$ 20$\pi$ rad/ns and $\tau=$ 5 ns. (c) $M_c$ as a function of the time delay, $\tau$. Other parameters are $\kappa=$ 30 ns$^{-1}$ and $\sigma=$ 20$\pi$ rad/ns. The simulations are averaged over 20 stochastic realizations of the initial conditions and frequency distribution.}
\end{figure}

One of the characteristic features of crowd synchronization is the scaling of the critical number
of elements with the degree of diversity in the population and with the coupling coefficient \cite{NatureBridge}. In our case diversity is caused by the different frequencies $\omega_j$ of the lasers. Figures~\ref{fig:3}(a,b) show the dependence of the critical number of lasers on the width $\sigma$ of the frequency distribution and on the coupling strength $\kappa$. The results show that
$M_c$ increases linearly with $\sigma$, while the dependence with $\kappa$ follows a power law with negative exponent, as occurs in \cite{NatureBridge}. 
As expected, the larger $\sigma$ the more different the lasers, and more lasers are required to generate the synchronized state. A broad frequency distribution leads to a reduction in the size of the clusters of lasers with similar $\omega_j$ showed in Fig. \ref{fig:1}(b), and a corresponding decrease in the coherent intensity.
On the other hand, the larger the coupling strength the smaller the minimum number of lasers required to synchronize the system [Fig. \ref{fig:3}(b)].

We have also considered the effect of the coupling delay $\tau$ on the transition to the synchronized state. As shown in Fig.~\ref{fig:3}(c), for short delays (compared with the characteristic time scales of the laser)
$M_c$ exhibits a sharp resonance at a $\tau$ corresponding to the cavity decay time, but for longer
delays $M_c$ is reduced and is almost independent of $\tau$.
When the coupling delays are not identical \cite{lee}, results similar to those of Fig.~\ref{fig:2} are found, but for larger coupling strengths. In that case the synchronized dynamics may be characterized by periodic fluctuations of small amplitude, or even
steady state emission.

In conclusion, we have shown numerically that a system of non-identical semiconductor lasers coupled to a common hub laser with time delay can be synchronized with zero lag. The transition to the synchronization occurs above a certain critical number $M_c$ of coupled lasers, provided the pump current of the hub laser is smaller than the solitary pump current threshold $\mu_{th}$. The type of synchronization transition can be controlled via the pump current of the star lasers: a gradual (second-order-like) transition is observed for star lasers with $\mu>\mu_{th}$, and an abrupt (first-order-like) transition arises for $\mu<\mu_{th}$. A similar behavior has been exhibited by a chemical quorum sensing system \cite{ScienceChemyOscil}.
The critical number of lasers increases linearly with the width of frequency distribution, and depends on the coupling
strength via a power-law with negative exponent, in agreement with the crowd synchronization transition reported in the Millenium bridge \cite{NatureBridge}. On the other hand, the coupling delay reduces the critical number of lasers while it has no influence on it for large enough time delays, even though the delay is evident through the lag time with which the hub laser is synchronized with the star lasers (which are synchronized isochronously to one another).

This research was supported in part by U.S. AFOSR grant FA9550-07-1-0238, the spanish MCI through project FIS2009-13360, and the AGAUR, Generalitat de Catalunya, through project 2009 SGR 1168. C.M. and J.G.O. are also supported by the ICREA Academia programme. R.R.'s research is partly supported by a DOD MURI grant (ONR N000140710734). R.R. thanks Nir Davidson and Asher Friesem of the Weizmann Institute (supported by a joint BSF grant) for ongoing research and discussions of laser phase-locking and synchronization. We also acknowledge E. Ott for his valuable comments and suggestions.


\begin{thebibliography}{99}
\bibitem{PumpAbove}  In the case that the hub laser is pumped above the solitary threshold, $\mu_H>\mu_{th}$, the synchronized state emerges progressively but without clustering, as a direct consequence of the intrinsic dynamics of the hub.
\end{thebibliography}
\end{document}